\documentclass[sts]{imsart}

\RequirePackage{amsthm,amsmath,amsfonts,amssymb}
\RequirePackage[numbers]{natbib}

\startlocaldefs

\usepackage[dvipsnames]{xcolor}
\usepackage[citecolor=CadetBlue, linkcolor=CadetBlue, colorlinks=true, urlcolor=CadetBlue]{hyperref}

\endlocaldefs

\begin{document}

\begin{frontmatter}

\title{Seven Principles for Rapid-Response Data Science: Lessons Learned from Covid-19 Forecasting}%\title{A sample article title with some additional note\thanksref{T1}}
\runtitle{Seven Principles for Rapid-Response Data Science}
%\thankstext{T1}{A sample of additional note to the title.}

\begin{aug}
% \author[A]{\fnms{Name1} \snm{Surname1}\thanksref{t1}\ead[label=e1]{first@somewhere.com}},
\author[A]{\fnms{Bin} \snm{Yu}\ead[label=e1]{binyu@berkeley.edu}}
\and
\author[B]{\fnms{Chandan} \snm{Singh}\ead[label=e2]{cs1@berkeley.edu}}

% \author[C]{\fnms{???} \snm{???}\ead[label=e3]{???@???}}
%%%%%%%%%%%%%%%%%%%%%%%%%%%%%%%%%%%%%%%%%%%%%%
%% Addresses                                %%
%%%%%%%%%%%%%%%%%%%%%%%%%%%%%%%%%%%%%%%%%%%%%%
\address[A]{Bin Yu is Professor in the Statistics Department and EECS Deparment as the University of California, Berkeley \printead{e1}.}
\address[B]{Chandan Singh is a PhD Candidate in the EECS Deparment as the University of California, Berkeley \printead{e2}.}
% \address[A]{???, \printead{e1}.}
% \address[B]{???, \printead{e2}.}
% \address[C]{???, \printead{e3}.}
\end{aug}

\begin{abstract}
In this article, we take a step back to distill seven principles out of our experience in the spring of 2020, when our 12-person rapid-response team used skills of data science and beyond to help distribute 340,000+ units of Covid PPE.
This process included tapping into domain knowledge of epidemiology and medical logistics chains,  curating a relevant data repository, developing models for short-term county-level death forecasting in the US, and building a website for sharing visualization (an automated AI machine).
The principles are described in the context of working with Response4Life, a then-new nonprofit organization, to illustrate their necessity. 
Many of these principles overlap with those in standard data-science teams, but an emphasis is put on dealing with problems that require rapid response, often resembling agile software development.
The technical work from this rapid response project resulted in a paper~\cite{altieri2021curating}; see also this interview for more background~\cite{yu2021interview}.\\
\end{abstract}


\end{frontmatter}
%%%%%%%%%%%%%%%%%%%%%%%%%%%%%%%%%%%%%%%%%%%%%%
%%%% Main text entry area:

\section{The decision to engage: preparedness and willingness} 
It was the evening of Friday, March 20, 2020, when one of the authors (Yu) saw an email from a colleague at Berkeley engineering asking for data-science help for a new nonprofit organization called Response4Life that aimed to distribute PPE (personal protection equipment) to Covid hot-spots in the US.
At this time, the burgeoning epidemic was taking hold in the US, and media outlets were crowded by reports on the huge shortage of PPEs such as masks, even for medical doctors and healthcare workers.

Within minutes, Yu replied to the email to get connected with Response4Life.
She also knew that her group, consisting of both statistics and EECS students/postdocs, was well-prepared to jump on this opportunity; they collectively held a research goal of solving real-world problems, possessed the necessary statistics/machine learning and computing skills, and had a collaborative culture in place, all tools necessary in order to move very fast as a team.
The group had good communication channels for resolving differences, a fair credit-sharing habit, and all members were already working together in overlapping teams of 2-5 people.

By Saturday, several group members had been contacted, including the other author (Singh), a PhD student from computer science. All signed up without hesitation for a warlike two-month engagement with Response4Life. Singh, a super-organizer, also agreed to be the deputy for the project with PI Yu.
In a few days, a team of 12 formed from other group members who joined voluntarily:
Nick Altieri (who delayed finishing his thesis), James Duncan,  Raaz Dwivedi, Karl  Kumbier (former group member),  Xiao Li, Robert Netzorg, Briton Park,  Yan Shuo  Tan,  Tiffany  Tang,  and Yu (Hue)  Wang.
 
After discussing with the Response4Life team, it became clear that this was a very unusual data-science project, with no data in hand or plans on how to find it.
Forecasts were needed to inform which hot-spot hospitals to send PPE, but this would require first finding relevant data, developing short-term (e.g. 5 days ahead) forecasting models, and integrating forecasts into a Response4Life Salesforce for use.
As a matter of fact, Response4Life was founded only a week before by Rick Brennan (founder of Airtime Aviation Holdings, LLC, engineer / entrepreneur, pilot) with mostly rotating short-term volunteers (e.g. with availability of two weeks).
Don Landwirth (angel investor, board member of Maker Nexus, advisor and mentor to dozens of entrepreneurs and executives) was the lead on connecting with makers and the Salesforce logistics platform.

\section{Effective human organization: divide and conquer} Since time was running very short, it was crucial to efficiently split up different aspects of the project amongst various people.
% for the team to split into subteams that handled different aspects of the project (and have short frequent meetings with larger groups to stay in sync).

Specifically, the team had two short weeks to come up with daily forecasts, at least 5 days ahead, for each of the 7000+ hospitals in the US and put them on the Salesforce platform for PPE distribution. With a team of 12 people, we naturally had to divide and conquer.
Subteams were formed: a data team (Tang and Wang), a modeling team (Altieri and Li), a logistics/visualization team (Singh and Duncan) to transport our prediction results to the Salesforce platform/website, and a PR team (Dwivedi, Netzorg, and Tan) to help organize many volunteers to Response4Life, scout media reports to qualitatively validate our predictions, and call hospitals to find out whether they would be interested in receiving PPE.

The teams were orchestrated following principles similar to agile development~\cite{cockburn2001agile}.
Short frequent large-group meetings were held to set goals for each team and evaluate progress.
These large meetings with the entire team's input helped to bridge the fast pace nature of the project with the more familiar academic environment of critique and iteration.
Google Docs was used as the primary software to organize the teams, with each team having a prioritized list of action items that were updated as goals evolved.

Our rapid-response team also met, first daily and later a couple of times a week, as a team and sub-teams to plan and sync-up at a very fast pace.
The authors attended daily Response4Life meetings (Yu for the 8 am leadership team meeting and Singh for the 3 pm logistics meeting).
After a couple of weeks, the meetings became less regular, but still multiple times a week.

\section{Gathering data and context: scraping, human contacts, and media reports}
Gathering data was an immediate challenge; it turned out that web scraping, human contacts, and monitoring media reports were all a crucial part of gathering data.

The first challenge was deciding which quantity to model.
We quickly found from media reports Covid-19 death-count data at the county level from USAFacts (and later NYT), but very little other information at the county level.
As a result, we decided to predict death-counts for each county.
We did not want to predict case numbers believing that (accumulated) death counts were more reliable and relevant to hospitalization than case numbers; this was particularly true at the beginning of the pandemic when case numbers were not uniformly reported.

Next, these county-level predictions were translated into an actionable hospital-level severity index.
At the time, hospital-level Covid-19 information was not available, so in order to impute hospital-level counts, we searched for information on individual hospitals through open sources, contacts and emails.

We then use the size of each hospital (measured by the total number of employees), to impute the death count for the hospital from each county, (relative to the total number of hospital employees in the county).
Then we combined features such as current (imputed) death counts, predicted (imputed) death counts, and ICU beds
% (??I don't believe we used ICU beds, did we??), % yes, we did
to assign each hospital a severity index, taking on values low, medium, and high.

Many other static relevant datasets with information on useful factors for predicting Covid-19 mortality were also collected, such as demographics and health risk factors.
Later on, more diverse dynamic data was added, such as social mobility data collected from many different sources such as Google and Apple maps, mask-wearing survey data, and safegraph social-distancing data.
This list continued to grow throughout the evolution of the pandemic to include data such as interventions that different counties/states made to spread gatherings.
We spoke to friends and colleagues who were practicing or studying medical doctors and researchers to understand what factors (health, social, economic) might impact death count in addition to past death counts.
Everyone we contacted was incredibly helpful.

\section{Data quality control: in-house data cleaning and curation}
Data quality control was critical for dealing with an influx of messy data.
It was important to vet the newly incoming data and set up a pipeline to easily adapt to unexpected changes from incurring data sources.
% ??we need reference the paper here its the data curation section??

The data team responded quickly to any lead from the team and contacts to scrape, clean, and curate before depositing in our repository (details in \cite{altieri2021curating}).
The dynamic data such as death and case counts were curated by different sources which used different data collection protocols, requiring cleaning for downstream tasks such as forecasting.
To efficiently gather and preprocess the data, the data team put an automatic scraping and cleaning procedure in place after some trial-and-error days.
This procedure required a large amount of maintenance, as the incoming data from different sources would sometimes change without warning (e.g. including previously-missed case counts or changing the code interface required to gather data).
Since the beginning of the project, many Ph.D. students in Yu-group teams, as well as external volunteers, helped with the data team in terms of finding useful datasets to be added as well as cleaning the data.
Eventually, two Ph.D. students kept on maintaining the pipeline and adding new datasets. Their skills of setting up periodic jobs on AWS EC2 instances, data cleaning via the Python Pandas package, and web-scraping were very handy.

Though information was desired at the hospital level, most data was available at different levels of granularity, e.g. death-counts at the county level and testing data at the state level.
Thus, human judgement calls were made on how best to impute between the different levels of granularity; for example, as mentioned earlier, county-level deaths were assigned to different hospitals proportional to each hospital's size.

\section{Speedy development and validation of many prediction algorithms}
After a quick examination of early death-counts, it became clear that standard epidemiological models were not yet doing a precise job at forecasting county-level Covid-19 deaths.
As a result, we turned to data-driven methods for forecasting, that would adapt to the data at hand.
With a pressing time demand and a rapidly evolving pandemic, we decided to build on past work of one author that weighs and combines multiple prediction schemes (with a forgetting factor) developed successfully for audio compression at Bell Labs 20 years ago~\cite{schuller2002perceptual}.
This is due to a recognition of the similar dynamic nature in audio and pandemic data and the proven dynamic adaptivity of the weighted combination strategy to good predictors.

% However, the basic predictors to combine had to differ from the audio predictors.
We first began by developing five extremely simple models to use as baselines for forecasting county-level deaths.
For each county, these predictors used the past time-series of 
% Five predictors were simultaneously developed by the modeling team using 
% past 
death and case counts, along with the demographic, social, economic, and health factors from the county and its neighboring counties.
In the end, trend-following, simple and transparent linear and exponential predictors stood out in terms of prediction performance (under three different reasonable loss functions) on future data (which arrives every day for 3000+ counties).
The best-predicting models used only past and neighboring county-level Covid death-counts to forecast future death-counts.

To improve upon these simple baseline models, we combined and weighted them (with a forgetting factor) to form the CLEP predictor in our paper~\cite{altieri2021curating}, where CLEP stands for Combined Linear and Exponential Predictor.
The weight on each predictor was higher if the predictor did a better job fitting recent data.
These weights predictors were interpretable because they corresponded well with the linear regime or the exponential regime that the pandemic fell in.
That is, in a linear regime of the pandemic, the linear predictor received more weight and vice versa.

The two members of the modeling team helped to implement different predictors in parallel. Familiarity with data analysis and python modeling packages such as \texttt{scikit-learn} \cite{pedregosa2011scikit}, \texttt{statsmodels} \cite{seabold2010statsmodels}, and \texttt{imodels} \cite{imodels2021} helped to develop models quickly and effectively.

\section{Uncertainty: measurement and empirical validation} 

Facing a dynamic future, it was important to assess uncertainty through a prediction interval with a justified level of confidence.

Prediction residuals are an obvious source to use for such an interval.
We used the prediction residuals of our model to defined \text{maximum absolute error prediction intervals} for predicted death-counts~\cite{altieri2021curating}.
Specifically, an interval was constructed using the maximum absolute relative error to add and subtract from the predicted (accumulated) death count of the future 5th (or 7th or 14th) day.
This construction can be seen as a form of generalized conformal analysis~\cite{vovk2005algorithmic}.
With a theoretical argument, we can see that the coverage of such intervals is around 80\%; this was more or less validated by comparing with the observed (accumulated) death counts (see~\cite{altieri2021curating} for details).
Additionally, these intervals require that errors are exchangeable across different days.
We restrict our interval construction to only use the previous five days; over this time period, the residuals empirically appeared exchangeable.

These intervals around our predictions would help inform the time needed for physical distribution of PPE from the makers (making sure the lower bound of the interval was larger than the last observed accumulated death count).
% With the right scale of relative error, the residuals appeared exchangeable after an empirical investigation by the modeling team.
The possibility of evaluating our predictions and intervals with ever coming new data every day was really the silver lining in an extremely challenging project and gave us empirically validated confidence in what we do.

\section{Communicating results: interactive visualizations, open-source code, and a web interface} Quickly and effectively communicating results was extremely important in our rapid-response setting. We did this through a combination of open-source code and visualizations put onto our website.

From the beginning, all our work was open-source on Github, making it easy for other groups to use our code and processed data.\footnote{\url{https://github.com/Yu-Group/covid19-severity-prediction}.}
Data was saved in both raw/processed form and updated daily, allowing other groups to make use of any subset of the data repository they found useful.
All data was placed into a common table-format which could then be easily distributed as a csv-file (or compressed table formats).
We also stored the daily-updated forecasts numerically on github, making them easy to access and compare against.

Moreover, we built interactive visualizations of the data, along with our forecasts, into a website using Github Pages\footnote{\url{https://covidseverity.com/}}.
These visualizations enabled much better exploration of the data, especially for specific counties and for comparing geographic areas.
Basic web skills, such as using HTML/Javascript were crucial for quickly setting up this website, and adding basic elements to it.
For actually producing the visualizations, various python visualization libraries such as Matplotlib and Plotly were useful, particularly their extensions for geographical-data visualizing.
Setting up compute on AWS allowed for easily automating daily updates to these visualizations as well as for hosting interactive visualizations which could not be simply uploaded to a static site.
The forecasts were also integrated into the Covid-19 Atlas\footnote{\url{https://theuscovidatlas.org/}} at the University of Chicago .

Three grades of severity (high, medium, and low) based on our 5-day predictions were put on the Salesforce platform
for PPE distribution within 18 days of our engagement, with non-stop long hours every day from core team members and people at Response4Life.
There were no other county-level prediction models available in the US until after our paper was submitted on May 16, 2020.
At the present date, we have taken down the realtime updates due to the availability of other county-level sites and our lack of sustainable resources for maintaining the website to deal with changing incoming data formats.
% at the source??? 

% \section{Building with maintenance in mind}: tests, 

% \section{Importance of good people, and combined skills of applied statistics and computer science}
\section*{Conclusion}
To conclude, this project was an intense and rewarding experience for everyone on our rapid-response data-science team, especially given the fact that our predictions helped inform the shipment of at least 349,000 face shields to doctors and healthcare workers (14,000 through Response4Life and the rest through Maker Nexus), at a time when they were direly needed.
It would not have been possible without the wonderful Response4Life people, our combined skills in applied statistics, machine learning, signal processing and coding, and awesome and timely support from friends, family, and colleagues.

%%%%%%%%%%%%%%%%%%%%%%%%%%%%%%%%%%%%%%%%%%%%%%
%% Single Appendix:                         %%
%%%%%%%%%%%%%%%%%%%%%%%%%%%%%%%%%%%%%%%%%%%%%%
%\begin{appendix}
%\section*{???}%% if no title is needed, leave empty \section*{}.
%\end{appendix}
%%%%%%%%%%%%%%%%%%%%%%%%%%%%%%%%%%%%%%%%%%%%%%
%% Multiple Appendixes:                     %%
%%%%%%%%%%%%%%%%%%%%%%%%%%%%%%%%%%%%%%%%%%%%%%
%\begin{appendix}
%\section{???}
%
%\section{???}
%
%\end{appendix}

%%%%%%%%%%%%%%%%%%%%%%%%%%%%%%%%%%%%%%%%%%%%%%
%% Support information, if any,             %%
%% should be provided in the                %%
%% Acknowledgements section.                %%
%%%%%%%%%%%%%%%%%%%%%%%%%%%%%%%%%%%%%%%%%%%%%%
\begin{acks}[Acknowledgments]
Thanks to Tiffany Tang, Hue Wang, Nick Altieri and Xiao Li for their helpful input to the paper, as well as two referees for their constructive comments that improved the paper.
\end{acks}
%%%%%%%%%%%%%%%%%%%%%%%%%%%%%%%%%%%%%%%%%%%%%%
%% Funding information, if any,             %%
%% should be provided in the                %%
%% funding section.                         %%
%%%%%%%%%%%%%%%%%%%%%%%%%%%%%%%%%%%%%%%%%%%%%%
\begin{funding}
Partial support of a grant from The Center for Information Technology Research in the Interest of Society and the Banatao Institute (CITRIS), University of California, is gratefully acknowledged.
\end{funding}

%%%%%%%%%%%%%%%%%%%%%%%%%%%%%%%%%%%%%%%%%%%%%%
%% Supplementary Material, including data   %%
%% sets and code, should be provided in     %%
%% {supplement} environment with title      %%
%% and short description. It cannot be      %%
%% available exclusively as external link.  %%
%% All Supplementary Material must be       %%
%% available to the reader on Project       %%
%% Euclid with the published article.       %%
%%%%%%%%%%%%%%%%%%%%%%%%%%%%%%%%%%%%%%%%%%%%%%
%\begin{supplement}
%\stitle{???}
%\sdescription{???.}
%\end{supplement}

%%  MR numbers will be added by VTeX.                      %%
%%                                                         %%
%%  Use \cite{...} to cite references in text.             %%
%%                                                         %%
%%%%%%%%%%%%%%%%%%%%%%%%%%%%%%%%%%%%%%%%%%%%%%%%%%%%%%%%%%%%%

%% if your bibliography is in bibtex format, uncomment commands:
\bibliographystyle{imsart-number} % Style BST file (imsart-number.bst or imsart-nameyear.bst)
% \bibliography{refs}       % Bibliography file (usually '*.bib')

%% or include bibliography directly:
% \begin{thebibliography}{}
% \bibitem{b1}
% \end{thebibliography}

\end{document}